\documentclass[preprint,groupedaddress,showpacs]{revtex4}%

\usepackage{graphicx}
\usepackage{amsmath}
\usepackage{epsfig}
\usepackage{color}

\graphicspath{{Pics/}}

\begin{document}

\title[Relativistic theory on ion reflection from electrostatic shocks]{Relativistic generalization of formation and ion reflection conditions in electrostatic shocks}

\author{A. Stockem\(^{1}\)}
\email[Electronic address: anne.stockem@ist.utl.pt]{}
\author{E. Boella\(^{1,2}\)}
\author{F. Fiuza\(^{1,3}\)}
\author{L. O. Silva\(^{1}\)}
\email[Electronic address: luis.silva@ist.utl.pt]{}
\affiliation{\(^{1}\)GoLP/Instituto de Plasmas e Fus\~ao Nuclear - Laborat\'orio Associado, Instituto Superior T\'ecnico, Lisboa, Portugal\\
\(^{2}\)Dipartimento Energia, Politecnico di Torino, Torino, Italy\\
\(^{3}\)Lawrence Livermore National Laboratory, California}

\date{\today}

\begin{abstract}
The theoretical model by Sorasio et al. (2006) for the steady state Mach number of electrostatic shocks formed in the interaction of two plasma slabs of arbitrary density and temperature is generalized for relativistic electron and non-relativistic ion temperatures. We find that the relativistic correction leads to lower Mach numbers, and as a consequence, ions are reflected with lower energies. The steady state bulk velocity of the downstream population is introduced as an additional parameter to describe the transition between the minimum and maximum Mach numbers in dependence of the initial density and temperature ratios. In order to transform the soliton-like solution in the upstream region into a shock, a population of reflected ions is considered and differences to a zero-ion temperature model are discussed. 
\end{abstract}


%\keywords{}

\maketitle

%%%%%%%%%%%%%%%%%%%%%%%%%%%%%%%%%%%%%%%%%%%%%%%%%%%%%%%%%%%%%%%%%%%%
%%%%%%%%%%%%%%%%%%%%%%%%%%%%%%%%%%%%%%%%%%%%%%%%%%%%%%%%%%%%%%%%%%%%
\section{Introduction}

Collisionless shocks can efficiently accelerate charged particles to high energies and its study is of interest to a wide range of scenarios, e.\,g.\ space and astrophysics, especially in the context of acceleration of cosmic rays to ultra high energies (up to \(\sim10^{21} \)\,eV) \cite{CB11}, laser-plasma interactions with applications in proton therapy of tumors \cite{Bulanov} or injection of particles for conventional accelerators \cite{Krushelnick}, and inertial confinement fusion \cite{Roth}. Electrostatic shocks have been observed in interplanetary space \cite{Lindqvist, Holback, Temerin}, in particular in the ionosphere \cite{Shelly} and in the auroral region \cite{Carlson}, where they arise from plasma cloud collisions at the interaction between solar wind and magnetosphere or between solar wind and interstellar medium in the heliosphere region \cite{Krimigis}.

Recently, electrostatic shocks have been propelled into the focus of research due to the ability of generating high Mach numbers in compact laboratory systems \cite{NP, PRL}, which could provide an alternative to the costly standard synchrotron accelerators. It was found that ions are efficiently accelerated in electrostatic shocks by reflection from the electrostatic potential with twice the shock velocity in the rest frame of the upstream population \cite{SM04}. Shocks with moderate Mach numbers were generated in a laboratory experiment, where ion beams with 20 MeV were produced \cite{NP}. The trend towards \(>250\)\,MeV/nucleon with a quasi-monoenergetic profile, which is relevant for the treatment of deep seated tumors \cite{FA06,MF08}, was demonstrated by particle-in-cell simulations of laser-driven shock acceleration \cite{PRL}.

The theoretical framework for the interaction of two plasma slabs leading to non-linear structures with large Mach numbers \cite{FS70,FF71} has been generalized by \citet[]{S06} for arbitrary plasma temperatures and densities. In this paper, we extend this work to relativistic electron and non-relativistic ion temperatures. The conditions for shock solutions are derived from an analysis of the Sagdeev potential \cite{Sagdeev-RPP-1966}, being fully described by the ratios of the initial densities and temperatures. A cold ion approximation allows for a determination of the condition for reflection and subsequent acceleration of ions by the electrostatic shock potential, which at the same time provides only the necessary condition for shock formation. The transition from such a soliton description, where no reflected or trapped ions are included, to a full shock solution is done with numerical techniques.

In section \ref{sec2} the theoretical framework is described and the necessary conditions for obtaining a shock-like solution are given. The dependence of the steady-state shock Mach number on the initial plasma temperature and density ratios is analyzed in section \ref{sec3}, and the underlying characteristics leading to a minimum and a maximum value for the Mach number are investigated, considering the impact of a downstream velocity on the shock Mach number. 
In section \ref{sec:ionreflection} the impact of reflected ions is addressed with a kinetic description of the ions.
%The conditions to obtain ion reflection and high energy ions are addressed in section \ref{sec:ionreflection} and dissipative effects that can create the oscillatory downstream potential are discussed in section \ref{sec4}.
The results are summarized in section \ref{sec5}.

%%%%%%%%%%%%%%%%%%%%%%%%%%%%%%%%%%%%%%%%%%%%%%%%%%%%%%%%%%%%%%%%%%%%
\section{Electrostatic shock solutions from arbitrary upstream and downstream densities and electron temperatures} \label{sec2}

\begin{figure}[ht!]
\begin{center}
\includegraphics[width=7cm]{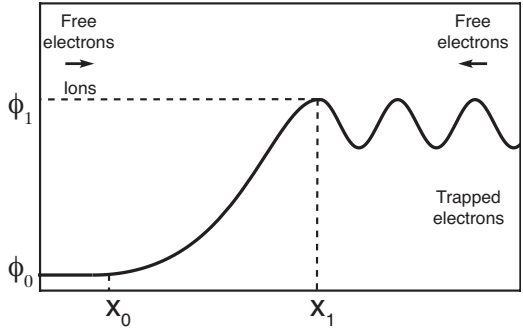}
\end{center}
\caption{Electrostatic shock formation from the interaction of free and trapped electrons and a population of cold ions. The electrostatic potential shows a monotonous increase from \(\phi_0\) in the upstream region (\(x< x_0\)) to \(\phi_1\) in the downstream region (\(x > x_1\)), where it becomes oscillatory.}\label{th}
\end{figure}

Plasmas composed of hot electrons and relatively cold ions are governed by ion sound waves with phase velocity \( \omega /k \simeq \sqrt{k_B T_e / m_i} = c_s\). The high ion inertia creates a restoring force to the thermally expanding electrons and wave steepening due to non-linear effects can lead to the generation of electrostatic shocks. 
The collision of two semi-infinite plasma slabs with arbitrary temperature and density provides the environment for electrostatic shock formation if the fluid velocity is small compared to the thermal velocity of the electrons. In the rest frame of the shock, the upstream population of electrons and ions is moving with $v_{sh}$ towards the shock. An electrostatic potential is formed due to the different inertia of the particles, which is steady in time in the shock rest frame, and increases monotonically from $\phi_0$ at $x=x_0$ in the far upstream until it reaches its maximum $\phi_1$ at $x=x_1$, where the transition to the oscillatory downstream region is defined (see Fig. \ref{th}). The double layer is then maintained by six populations. Free electron and ion populations in the upstream and downstream regions, with kinetic energies higher than the potential energy, a population of trapped electrons whose kinetic energy is less than the potential energy and a population of reflected ions in the upstream region.

We make use of the Sagdeev formalism \cite{Sagdeev-RPP-1966} in order to determine the electrostatic potential. In this section, we focus on the electron kinetics in the upstream region, assuming zero ion temperature, and determine the condition for ion reflection. The impact of a population of reflected ions with non-zero temperature is addressed in section \ref{sec:ionreflection} together with a full description of the shock potential.

%%%%%%%%%%%%%%%%%%%%%%%%%%%%%%%%%%%%%
%%%%%%%%%%%%%%%%%%%%%%%%%%%%%%%%%%%%%%%

\begin{figure}[b]
\begin{center}
\includegraphics[width=8cm]{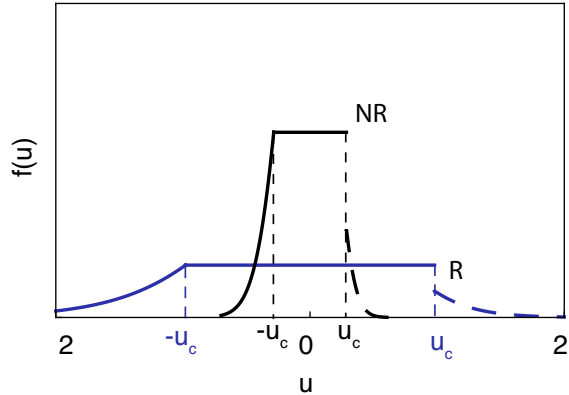}
\end{center}
\vspace{-12pt}
\caption{Electron distribution functions upstream (\(f_0\) -- dashed) and downstream (\(f_1 + f_{1t}\) -- solid) for relativistic electron temperature (R) \(\mu_0 = 5\) (blue) and non-relativistic temperature (NR) \(\mu_0 = 50\) (black) with \(\Gamma = 3\), \(\Theta = 2\), \(e (\phi_1 - \phi_0)/m_e c^2 = 2\) and \(\beta_{sh} = 0.02\).}\label{distr_fun}
\end{figure}

The electron populations are treated kinetically, while the ions are treated as a fluid. The electron distributions have to be solution of the stationary Vlasov equation and can be determined knowing the particle distributions in the unperturbed plasma. The free (left) electron population, propagating from the upstream to the downstream region, is described by the relativistic drifting 1D J\"uttner distribution function \(  f_0(\gamma_0) = N_0 \mathrm K_1^{-1}(\mu_0) \, \gamma_0(\gamma_0^2-1)^{-1/2} \exp \! \left[ -\mu_0 \gamma_0 (1- \beta_0 \beta_{sh}) \right]\) \cite{LS10} with the normalized velocity of the electrons \(\beta_0 = v_0 / c >0\), the Lorentz factor \(\gamma_0 = (1- \beta_0^2)^{-1/2}\), the thermal parameter $\mu_0 = m_e c^2/ k_B T_0$, where \(T_0\) is the electron temperature, $m_e$ the electron mass and $k_B$ the Boltzmann constant, and normalized fluid velocity $\beta_{sh} = v_{sh}/c$. The normalization constant contains the density of the left electron population \(N_0\) in the upstream region \((x < x_0)\) and the modified Bessel function of the second kind \(K_1\). In the limit of nonrelativistic electron temperatures, \( \mu_0 \gg 1\), the distribution function is approximated by a Maxwell-Boltzmann distribution and the thermal velocity can be introduced as $v_{th,0} = \sqrt{k_BT_0/m_e} = c/ \sqrt{ \mu_0}$.
The free electrons in the downstream region (right population) are described by \( f_1(\gamma_1) =  N_1 \mathrm K_1^{-1}(\mu_1) \,  \gamma_1 (\gamma_1^2-1)^{-1/2} \exp \! \left[  -\mu_1 \gamma_1 (1- \beta_1 \beta_{d})  +  \frac{e (\phi_1 - \phi_0)}{k_B T_1}  \right] \) with \(\beta_1 = v_1/ c < 0\), where the parameters have the same meaning as for the left population and are indicated with a subscript 1, and \(\beta_d\) is the fluid velocity of the downstream. To facilitate comparison with the nonrelativistic model we stick to the notation of Sorasio et al. \cite[]{S06} and have multiplied the distribution function by a factor containing the potential difference \(\phi_1-\phi_0\).  $N_1$ represents the density of the right electron population in the far upstream and sums up with the density of the left population to the ion density \(N_i = N_0 + N_1\) at $x < x_0$ to guarantee charge neutrality. The trapped electrons are represented by the flat-top distribution function in the relativistic notation $f_{1t}= N_1 \mathrm K_1^{-1}(\mu_1) \,  \gamma_1 (\gamma_1^2-1)^{-1/2}  \exp (-\mu_1) $, according to the so-called ``maximum-density-trapping" approximation \cite{Schamel-PP-1972, Montgomery-PP-1969}, which guarantees $f_1(\gamma_1=\gamma_c)=f_{1t}$ at the critical Lorentz factor $\gamma_c= 1 + e (\phi_1 - \phi_0) / m_e c^2 $ that discriminates between free ($\beta_1<-\beta_{c}$) and trapped electrons ($|\beta_1|<\beta_{c}$), shown in Fig. \ref{distr_fun} where the more convenient parameter \(u= \beta \gamma\) has been introduced. Since the fluid velocities \(\beta_{sh}\) and \(\beta_{d}\) are small compared to the thermal velocities, we will neglect this dependence in the following calculations \cite{jay,Sorasio}.

%%%%%%%%%%%%%%%%%%%%%%%%%%%%%%%%%%%%%
%%%%%%%%%%%%%%%%%%%%%%%%%%%%%%%%%%%%%

We introduce the electron Lorentz factor $\gamma_e$ which accounts for the electrostatic potential in the shock frame and make use of the conservation of energy to write the upstream and downstream Lorentz factors as $\gamma_e= \gamma_0  + e (\phi - \phi_0) / m_e c^2 =  \gamma_1 - e (\phi_1 - \phi) / m_e c^2 \geq 1$. The electron density can then be computed as \(n_e =\int_{1}^{\infty}  f_e(\gamma_e) \, d \gamma_e \), obtaining the electron densities in the upstream region 
$n_0 (\Delta \varphi)=N_0  \, \mathrm K_1^{-1}(\mu_0)  \mathrm{e}^{\Delta \varphi}  \int_{1+ \Delta \varphi / \mu_0}^\infty \mathrm e^{-\mu_0 \gamma_e} \gamma_e (\gamma_e^2 - 1)^{-1/2} \, d \gamma_e$ and downstream region
$n_1 (\Delta \varphi)= N_0 \Gamma  \mathrm K_1^{-1}(\mu_0 / \Theta) \left[  \mathrm{e}^{\Delta \varphi / \Theta}  \int_{1+\Delta \varphi / \mu_0}^\infty \mathrm e^{-\mu_0 \gamma_e/\Theta}\gamma_e (\gamma_e^2 - 1)^{-1/2}  \, d \gamma_e+ 2  \mathrm e^{-\mu_0/\Theta} \sqrt{\left( 1+ \Delta \varphi/\mu_0\right)^2 -1 } \right]$ with the dimensionless quantities $\Delta \varphi =e (\phi-\phi_0) \mu_0 / m_e c^2$,  $\Gamma=N_1/N_0$ and $\Theta=\mu_0 / \mu_1$. 

Using the fluid equations for ion mass and energy conservation and assuming that the ions are cold and that none of them is reflected at the potential, the ion density can be determined as $n_i (\Delta \varphi)= N_i /\sqrt {1- 2 \, \Delta \varphi / M^2} $, where $M=v_{sh}/c_s$ is the ion acoustic Mach number, $c_s=\sqrt{{k_B T_0}/{m_i}} $ the upstream ion sound speed and $m_i$ is the ion mass. The particle densities are then combined in Poisson's equation
\begin{equation}\label{poisson}
\frac{\mathrm{d}^2 \Delta \varphi}{\mathrm{d} \chi^2} = - \frac{n_i(\Delta \varphi)}{N_i} + \frac{n_0(\Delta \varphi) + n_1(\Delta \varphi) }{N_i}, 
\end{equation}
where the normalized quantity $\chi={x}/{\lambda_D}$ with $\lambda_D=\sqrt{K_B T_0/(4 \pi e^2 N_i)}$ has been introduced.  Defining the right hand side of equation \eqref{poisson} as $-\mathrm{d}\Psi(\Delta \varphi)/\mathrm{d} \Delta \varphi$, the similarity to the harmonic oscillator can be immediately noticed and equation \eqref{poisson} can be seen as the motion of a pseudo-particle in the Sagdeev potential $\Psi(\Delta \varphi)$   \cite{Sagdeev-RPP-1966}, allowing to identify bounded solutions as possible shock solutions. 
Integration of equation \eqref{poisson} with respect to $\Delta \varphi$ leads to 
\begin{equation}
\frac{1}{2} \left( \frac{\mathrm{d} \Delta \varphi}{\mathrm{d} \chi} \right)^2+\Psi(\Delta \varphi)= const =:\Psi_0 \label{int_dphi}
\end{equation}
with \(\Psi_0 = \Psi(\varphi_0,M,\Gamma,\Theta)\) and the non-linear Sagdeev potential given by
\begin{equation} \label{Sag}
\Psi(\Delta \varphi,M,\Gamma,\Theta) = P_i(\Delta\varphi, M) - P_{e0}(\Delta \varphi, \Gamma) -P_{e1} (\Delta \varphi, \Gamma, \Theta)
\end{equation}
where the quantities $P_i$, $P_{e0}$ and $P_{e1}$ represent the ion, the upstream electron and the downstream electron pressures, respectively, which are defined as
\begin{eqnarray}
P_i(\Delta\varphi, M) &=&  M^2 \left(1- \sqrt{1-\frac{2 \Delta \varphi}{M^2}}       \right) \label{ionpressure}   \\
P_{e0}(\Delta \varphi, \Gamma, \mu_0) &=& \frac{1}{1+\Gamma} \left[  \frac{\mu_0}{K_1(\mu_0)} \int_1^\infty d \gamma \, \mathrm{e}^{-\mu_0 \gamma} \sqrt{\left( \gamma + \frac{\Delta \varphi}{\mu_0}   \right)^2 -1 } -1 \right]  \label{uppress}\\
P_{e1} (\Delta \varphi, \Gamma, \Theta, \mu_0) &=& \frac{\Gamma \Theta}{1+\Gamma} \left[  \frac{\mu_0 \mathrm e^{-\mu_0/\Theta} }{\Theta K_1(\mu_0/\Theta)} \left\{ \int_1^\infty d \gamma \, \mathrm{e}^{-\mu_0 (\gamma-1)/\Theta} \sqrt{\left( \gamma + \frac{\Delta \varphi}{\mu_0}   \right)^2 -1 }  \right. \right. \nonumber\\
&&+ \left. \left. s\, \sqrt{s^2-1} - \log \left[ s+\sqrt{s^2-1} \right]  \right\} -1 \right] \label{downpress}
\end{eqnarray}
with \(s = 1+ \Delta \varphi / \mu_0 \). In the case of nonrelativistic temperatures, \(\mu_0 \gg 1\), equations (\ref{uppress}) and (\ref{downpress}) can be integrated analytically, retrieving the expressions of \cite{S06}
\begin{eqnarray}
P_{e0}^{nr}(\Delta \varphi, \Gamma) &=& \frac{1}{1+\Gamma} \left( \frac{2 \sqrt{\Delta \varphi}} {\sqrt{\pi}} + \mathrm{e}^{\Delta \varphi} \operatorname{erfc} {\sqrt{\Delta \varphi}} -1 \right)   \label{uppressnonrel} \\
P_{e1}^{nr}(\Delta \varphi, \Gamma, \Theta) &=& \frac{\Theta \Gamma}{1+\Gamma}  \left( \frac{2}{\sqrt{\pi}} \sqrt{\frac{\Delta \varphi}{\Theta}} + \mathrm{e}^{\frac{\Delta \varphi}{\Theta}} \operatorname{erfc} \sqrt{ \frac{\Delta \varphi}{\Theta}} +\frac{8}{3 \sqrt{\pi}} \Delta \varphi \sqrt{\frac{\Delta \varphi}{\Theta^3}} -1\right)
\end{eqnarray}
with erfc the complimentary error function. Note that the explicit dependence on the upstream and downstream temperatures vanishes in this approximation. For highly relativistic electron temperatures, \(\mu_0 \ll 1 \), the pressures are approximated by
\begin{eqnarray}
P_{e0}^{r}(\Delta \varphi, \Gamma) &=& \frac{ \Delta \varphi (1-\mu_0) }{1+\Gamma} \\
P_{e1}^{r} (\Delta \varphi, \Gamma, \Theta) &=& \frac{ \Delta \varphi \Gamma}{\Theta (1+\Gamma) } \left[  \Delta \varphi \left( 1- \frac{\mu_0}{\Theta}\right) + \Theta  \left( 1+ \frac{\mu_0}{\Theta}\right)  \right]   \label{downpresshighrel}
\end{eqnarray}
and the explicit dependence on the temperatures is still maintained.
%%%%%%%%%%%%%%%%%%%%
\begin{figure}[htb]
\begin{center}
\includegraphics[width=7cm]{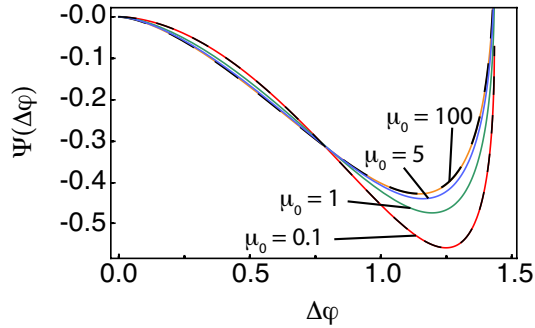}
\end{center}
\vspace{-12pt}
\caption{Sagdeev potential \(\Psi (\varphi) \) obtained from equations (\ref{Sag})-(\ref{downpress}) for \(M = 1.7\), \(\Gamma = 3\), \( \Theta = 1\) and \(\mu_0 = 0.1\) (red), 1 (green), 5 (blue), 100 (orange). Represented by dashed lines are the non-relativistic and highly relativistic approximations, given by equations (\ref{uppressnonrel})-(\ref{downpresshighrel}).}\label{fig:Sagdeev}
\end{figure}
%%%%%%%%%%%%%%%%%%%%
 Figure \ref{fig:Sagdeev} shows the Sagdeev potential, given by equations  (\ref{Sag})-(\ref{downpress}), for upstream electron temperatures \(\mu_0 = 0.1-100\) for a constant Mach number \(M\) and a comparison with the nonrelativistic and highly relativistic approximations (\ref{uppressnonrel})-(\ref{downpresshighrel}). A higher temperature leads to larger absolute values of the minimum of \(\Psi\), which results in lower values for the Mach number at which ion reflection sets in, as will be shown in the following section.

\section{Mach number dependence on initial parameters} \label{sec3}

\begin{figure}[ht!]
\begin{center}
\includegraphics[width=7cm]{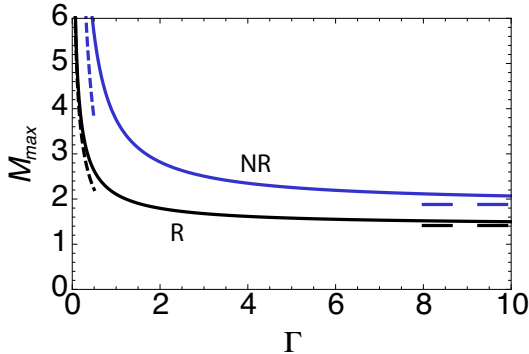}
\end{center}
\vspace{-12pt}
\caption{Maximum Mach number versus density ratio for the highly relativistic case (R) \(\mu_0 = 0.1\) (black) given by Eq.\ (\ref{Mmax_rel}) and the non-relativistic case (NR) given in \cite{S06} (blue) for a temperature ratio \(\Theta = 1\). The dependences for very small and very large density ratios are indicated by the dashed lines.}\label{fig:mmaxgamma}
\end{figure}

As can be seen from equation (\ref{ionpressure}), the model holds for \(\Delta \varphi < M^2/2 :=\Delta \varphi_{cr}\). The ion pressure becomes imaginary when the electrostatic potential exceeds the ion kinetic energy  
\begin{equation}
	e \Delta \phi > \frac{1}{2} m_i v_i^2 
\end{equation}
and the ions are reflected by the shock potential. We define the Mach number at which ion reflection sets in as the maximum Mach number \(M_{max}\). In order to determine possible shock solutions with \(M_{max}\), we use equation \eqref{int_dphi} which gives the condition for the existence of a monotonic double layer solution as $\tilde \Psi := \Psi(\Delta \varphi, M, \Gamma, \Theta)-\Psi_0<0$. For a given Mach number $M$, a soliton-like solution is possible only if the electron pressure exceeds the ion pressure.  The solutions are found numerically by solving \(\tilde \Psi(M^2/2, M, \Gamma, \Theta) =0\) and are shown in figure \ref{fig:mmaxgamma}.

%If \(\Psi_0 = \Psi(\varphi_0, M, \Gamma, \Theta) \not = 0\), the \(\varphi_0\) dependence is eliminated by expressing the Mach number at the local maximum \(\varphi_0\) by
%
%\begin{equation}\label{Mext}
%	M_{ext}(\varphi_0)= \sqrt{2 \varphi_0} /\left\{  1- \frac{ (1+\Gamma)^2}{\left[ e^{\varphi_0} \textrm{erfc} [\sqrt{\varphi_0}] + \Gamma  \left( 4\sqrt{\frac{\varphi_0}{\pi \Theta}} + e^{\varphi_0/ \Theta} \textrm{erfc} [\sqrt{\varphi_0/\Theta}] \right) \right]^2}  \right\}^{1/2}.
%\end{equation}
%
%The deviation to the approximation \(\varphi_0 =0\) is negligible because \(\Delta \varphi_{cr} \gg \varphi_{0} \).  

%----------------------------------------------------------------------------
%----------------------------------------------------------------------------

As already found in \cite{S06}, the analytical dependence of the maximum Mach number in the non-relativistic approximation is given by \(M_{max} = 3 \sqrt{\pi\Theta/8} (1 + \Gamma)/\Gamma \), which is \(M_{max} \approx 3 \sqrt{\pi\Theta/8}\) for large density ratios and has a \(M_{max} \propto \Gamma^{-1}\) dependence for low density ratios. Here, we find for the case of highly relativistic temperatures, \(\mu_0 \ll 1\), 
\begin{equation}\label{Mmax_rel}
	M_{max} = \sqrt{2 \Theta \left( 1 + \frac{1+ \mu_0}{\Gamma (1- \mu_0 / \Theta)} \right)},
\end{equation} 
which is displayed in figure \ref{fig:mmaxgamma} together with the non-relativistic expression. It can be easily seen that the maximum Mach number is constant for high density ratios as in the non-relativistic case, \(M_{max} \approx \sqrt{2 \Theta}\) for \(\Gamma \gg 1\), and has a dependence \(M_{max} \approx \sqrt{2 \Theta (1+\mu_0)/\Gamma (1-\mu_0 / \Theta)} \propto \Gamma^{-1/2} \) for \(\Gamma \ll 1\). The comparison of the non-relativistic and highly relativistic cases in figure \ref{fig:mmaxgamma} for a temperature ratio \(\Theta =1\) shows that for higher upstream electron temperatures the maximum Mach number is reduced, in agreement with the model for equal density and temperature ratios \cite{jay}.

%----------------------------------------------------------------------------
\begin{figure}[ht!]
\begin{center}
\includegraphics[width=7cm]{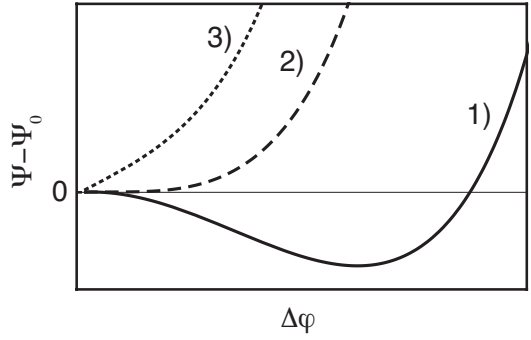}
\end{center}
\vspace{-12pt}
\caption{Different types (1) -- (3) of the Sagdeev potential \(\tilde \Psi (\Delta \varphi) = \Psi-\Psi_0\) with explanations of the  cases given in the text.}\label{fig:3types}
\end{figure}
%----------------------------------------------------------------------------

\begin{figure}[ht!]
\begin{center}
\includegraphics[width=7cm]{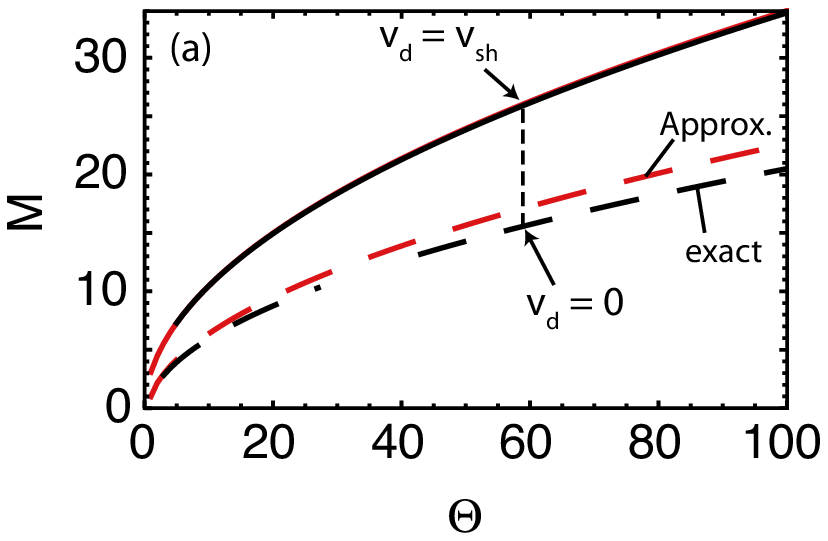}
\includegraphics[width=7cm]{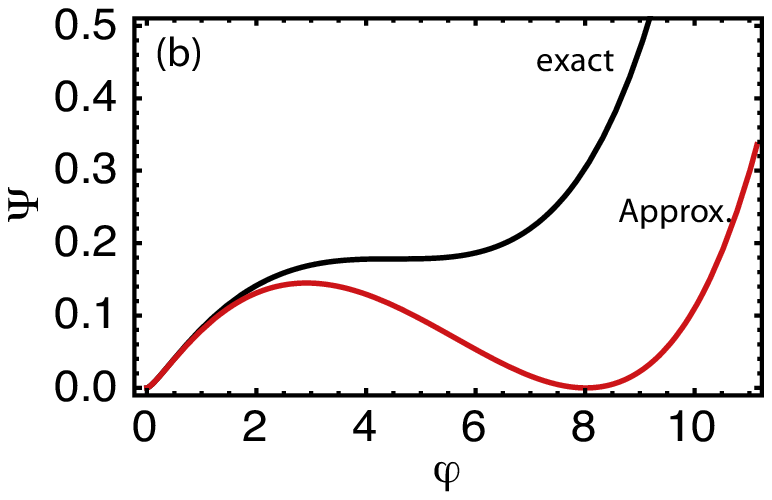}
\end{center}
\vspace{-12pt}
\caption{(a) \(M_{min}\) (dashed) and \(M_{max}\) (solid) for the approximation \(\varphi_0 = 0\) (red) and comparison with the exact solution (black) for \(\Gamma = 1\). The dashed vertical line shows the transition from \(M_{min}\) to \(M_{max}\) for downstream velocities \(0 \leq v_d \leq v_{sh} \). (b) Sagdeev potential for the minimum Mach number with \(\Delta \varphi = 0\)  (black) and comparison with the approximation \(\varphi_0=0\) (red).}\label{CompSorasio}
\end{figure}

We analyze now the lower limit and the range of possible Mach numbers for given temperature and density ratios. The shape of the Sagdeev potential and thus the existence of shock solutions depends on the choice of \(\Gamma\) and \(\Theta\) and we can distinguish three different types of solutions which are shown in Figure \ref{fig:3types}. Case (1) represents the case where shock solutions exist for $\tilde \Psi = \Psi-\Psi_0<0$ and  \(\Delta \varphi > 0\). While the monotonously growing Sagdeev potential in case (3) does not allow for shock solutions, case (2) defines the threshold with \(\Delta \varphi = 0\) and provides the conditions to determine the minimum Mach number, which are given by \( d \tilde \Psi / d \Delta \varphi  = 0 \) and \( \tilde \Psi (\Delta \varphi) = 0\). While in the highly relativistic limit \(M= 1\) is the lower limit, in the non-relativistic case a lower limit \(M>1 \) exists. 
The Sagdeev potential is expanded for \(\Delta \varphi \ll 1\) since we are looking for solutions \(\Delta \varphi \rightarrow 0\), obtaining
\begin{eqnarray}\label{approxPsi}
	\tilde \Psi(\Delta \varphi,M, \Gamma,\Theta) &\approx& \Delta \varphi^2 \left[ \frac{1}{2 M^2 ( 1- \frac{2 \varphi_0}{M^2})^{3/2}}  \right.\nonumber\\
	&+& \left. \frac{1}{2(1+\Gamma)} \left( \frac{1- \Gamma \Theta^{-1/2}}{ \sqrt{\varphi_0 \pi}} - e^{\varphi_0} \textrm{erfc} (\sqrt {\varphi_0}) - \frac{\Gamma}{\Theta} e^{\varphi_0/ \Theta} \textrm{erfc} (\sqrt {\frac{\varphi_0}{\Theta}}) \right) \right],
\end{eqnarray}
which is a function of the upstream potential \(\varphi_0\). The minimum Mach number can then be found by solving \(\tilde \Psi (\Delta \varphi,M(\varphi_0), \Gamma, \Theta) = 0 \) with the Mach number at the minimum of the Sagdeev potential given by
\begin{equation}\label{Mext}
	M(\varphi_0)= \sqrt{2 \varphi_0} /\left\{  1- \frac{ (1+\Gamma)^2}{\left[ e^{\varphi_0} \textrm{erfc} [\sqrt{\varphi_0}] + \Gamma  \left( 4\sqrt{\frac{\varphi_0}{\pi \Theta}} + e^{\varphi_0/ \Theta} \textrm{erfc} [\sqrt{\varphi_0/\Theta}] \right) \right]^2}  \right\}^{1/2}.
\end{equation}
For large temperature ratios we find small deviations from the approximation \(\varphi_0 = 0\) \cite{S06}, which is equivalent to \(\Psi_0 =0\) in equation (\ref{int_dphi}), see Figure \ref{CompSorasio} (a). Panel (b) shows the respective Sagdeev potentials with \(\Psi_0 = 0\) for the minimum Mach number according to \(\Delta \varphi = 0\) in black and for the approximated model with \(\varphi_0 = 0\) in red. The exact solution allows for the formation of electrostatic shocks at slightly lower Mach numbers.

%----------------------------------------------------------------------------
%----------------------------------------------------------------------------

The transition between minimum and maximum Mach number can be described as a function of the steady state ion speed in the downstream region, \(v_d\). In the rest frame of the shock, the upstream ions propagate towards the shock with velocity \(v_{i,us} := v_{sh}\) and are decelerated by the shock potential \(\varphi\) to velocities in the downstream region \(0 \leq v_{i,ds}  \leq v_{sh}\). The velocity is \(v_{i,ds} = 0\) if the ions are completely stopped by the potential and \(v_{i,ds} = v_{sh}\) if they are unaffected and stream freely in the downstream region. In the upstream frame this corresponds to ion downstream velocities \( -v_i \leq v_{i,du}:= - v_d \leq 0\). Starting once more from the energy conservation for ions, we can relate the downstream ion speed directly with the shock potential \(v_{d}/ c_s = M-\sqrt{M^2-2 \Delta \varphi} \) and use
\begin{equation}
	\Delta \varphi_d := \frac{M^2}{2} \left[ 1-\left( 1- \frac{v_d}{v_{sh}} \right)^2 \right]
\end{equation}
to find the zeros of the Sagdeev potential \(\tilde \Psi(\Delta \varphi_d ,M, \Gamma,\Theta)\) to determine the Mach number \(M\). This transition is shown in figure \ref{CompSorasio}a. When the shock propagates with a speed slightly above the minimum Mach number, the downstream population will have almost the same speed as the upstream population due to the small potential jump that has only a weak effect on the particles. At the maximum Mach number, the potential jump is so strong that the downstream population propagates with the same speed as the shock front.

%----------------------------------------------------------------------------
%\begin{figure}[ht!]
%\begin{center}
%\includegraphics[width=7cm]{Mofvd}
%\includegraphics[width=7cm]{MofY}
%\end{center}
%\vspace{-12pt}
%\caption{Mach number versus downstream speed with \(\Gamma = 1\), \(\Theta = 1.5\) (dashed),  \(\Gamma = 1\), \(\Theta = 1\) (solid),  \(\Gamma = 3\), \(\Theta = 1\) (dotted). %(b) Mach number versus density ratio for \(\Theta=1\) and \(v_d /v_{sh} = 1\) (solid), 0.8 (dashed) and 0.6 (dotted).
%}\label{fig:M_vd}
%\end{figure}
%----------------------------------------------------------------------------

%Figure \ref{fig:M_vd} shows that the Mach number is decreased if \(v_d < v_{sh}\) and the values for \(v_d = 0\) and \(v_d=v_{sh}\) correspond to the cases of the minimum and maximum Mach numbers. There are also included solutions with \(M<1\), which correspond to periodic waves \cite{BP70} and do not represent actual shock solutions.

\section{Ion reflection}\label{sec:ionreflection}

\begin{figure}[ht!]
\begin{center}
\includegraphics[width=6.8cm]{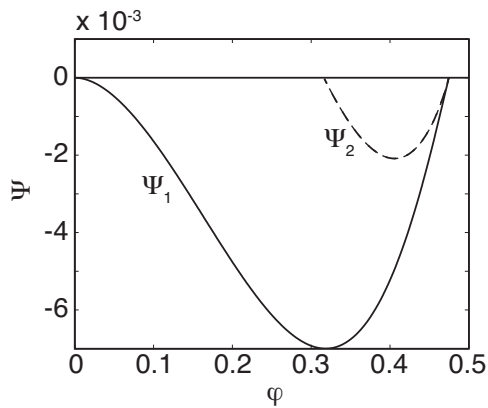}
\end{center}
\vspace{-24pt}
\caption{Sagdeev potentials for \(T_e = 10\) keV,  \(T_i = 0.5\) keV, \(M = 1.5\) and \(\Theta = \Gamma = 1\). Potential \(\Psi_1\) (solid) corresponds to \(0 \leq \chi \leq \chi_m\) and \(\Psi_2\) (dashed) to \(\chi > \chi_m \).}\label{fig7}
\end{figure}

So far, we have described the solitary solution in the upstream region and neglected the processes leading to a shock solution. A shock solution can arise due to different physical mechanisms that break the symmetry \cite{TK71}. For instance, a very small ion temperature is sufficient to lead to an oscillating solution (cp. \cite{Cairns}). To describe this, a population of reflected ions is included in the model. The electrostatic potential in the upstream region is computed as in section \ref{sec2} with the extension of a kinetic treatment of the ions. On the basis of \cite{B82,SB83,B86}, the ion populations are described by a Maxwellian distribution \(f_{i} =  \frac{n_i}{\sqrt{2\pi} v_{th,i}} \exp \left[ -\frac{1}{2 v_{th,i}^2} \left( \sqrt{v^2+ 2 c_{s0}^2\varphi} - c_{s0} M \right)^2 \right]\) with thermal velocity \(v_{th,i} = \sqrt{k_B T_{i}/m_i}\), which guarantees charge neutrality with the electrons in the far upstream region \(\chi \rightarrow -\infty\). The free particle population has velocities \(v > v_c = \sqrt{2 (\varphi-\varphi_1) c_{s0}^2}\) and the reflected population \(0 \leq v \leq v_c\). Since an exact analytical solution cannot be found, we solve the equations numerically. The Sagdeev potential \(\Psi_1\) is computed for \(\chi \leq \chi_m\) with \(\chi_m\) the position of the maximum of the electrostatic potential, where the connection is made to the oscillatory downstream region of the shock, which is described by a second Sagdeev potential \(\Psi_2\) (see figure \ref{fig7}). For the computation of the latter, two populations of free ions and electrons, as well as trapped electrons are considered. 
%----------------------------------------------------------------------------
%----------------------------------------------------------------------------

\begin{figure}[ht!]
\begin{center}
\includegraphics[width=6.8cm]{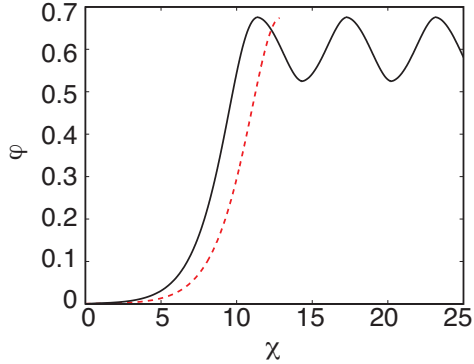}
\end{center}
\vspace{-24pt}
\caption{Electrostatic potential for \(T_e = 10\) keV, \(M = 1.62\), \(\Theta = \Gamma = 1\) and \(T_i = 0.5\) keV (solid black), \(T_i = 0\) (red dashed).}\label{fig8}
\end{figure}

Figure \ref{fig8} shows the corresponding electrostatic potential against the spatial coordinate, which consists of a monotonously increasing part for \(\chi \leq \chi_m\) and an oscillatory downstream region for \(\chi > \chi_m \). We also compare the solution where ion reflection was neglected (dashed red) with the extended model. For an ion temperature corresponding to 0.5 keV, we observe only a small deviation from the cold model. For the same potential difference, the maximum Mach number increases as it was expected \cite{S06}. 
%----------------------------------------------------------------------------
%----------------------------------------------------------------------------

\begin{figure}[ht!]
\begin{center}
\includegraphics[width=6.8cm]{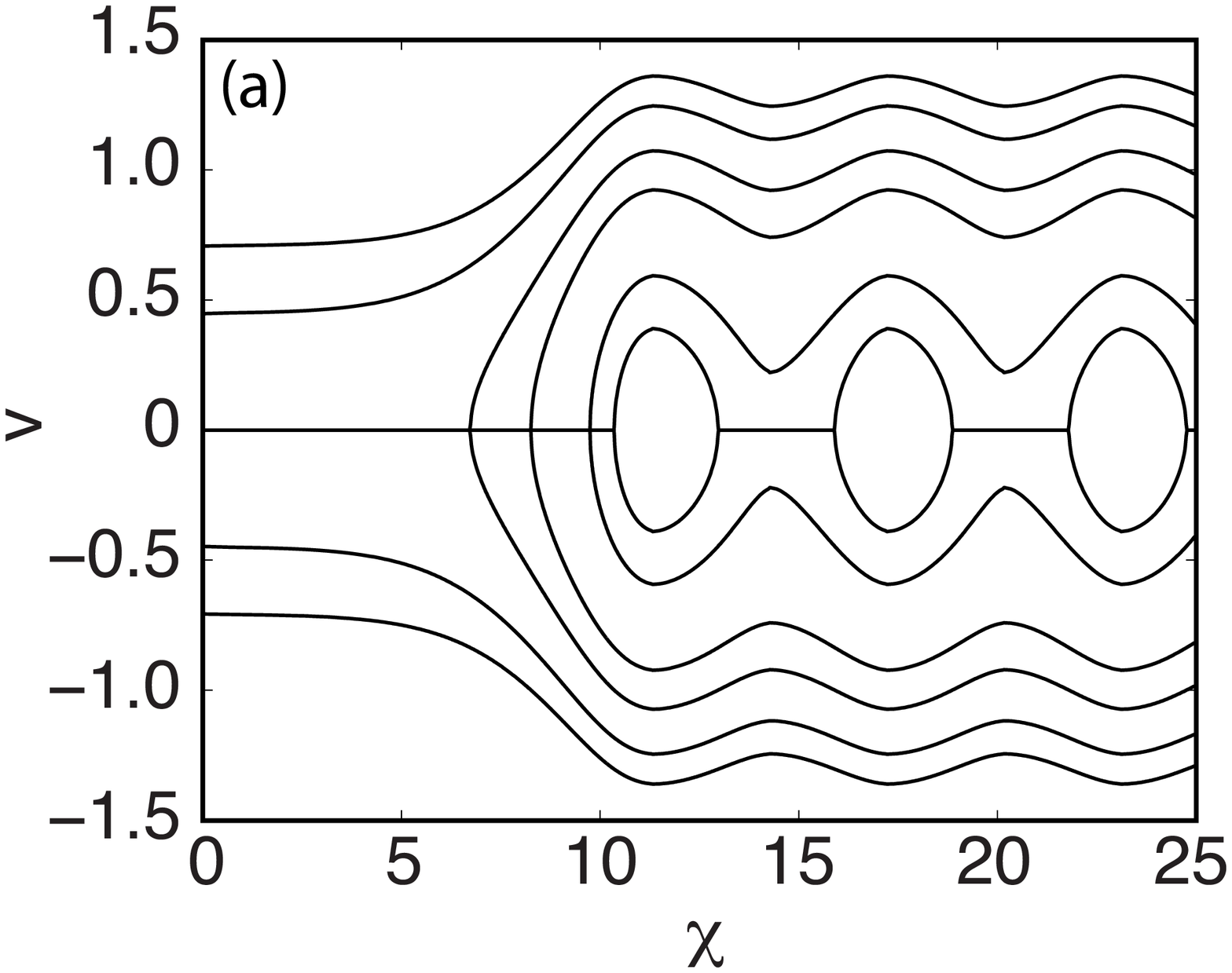}
\includegraphics[width=6.8cm]{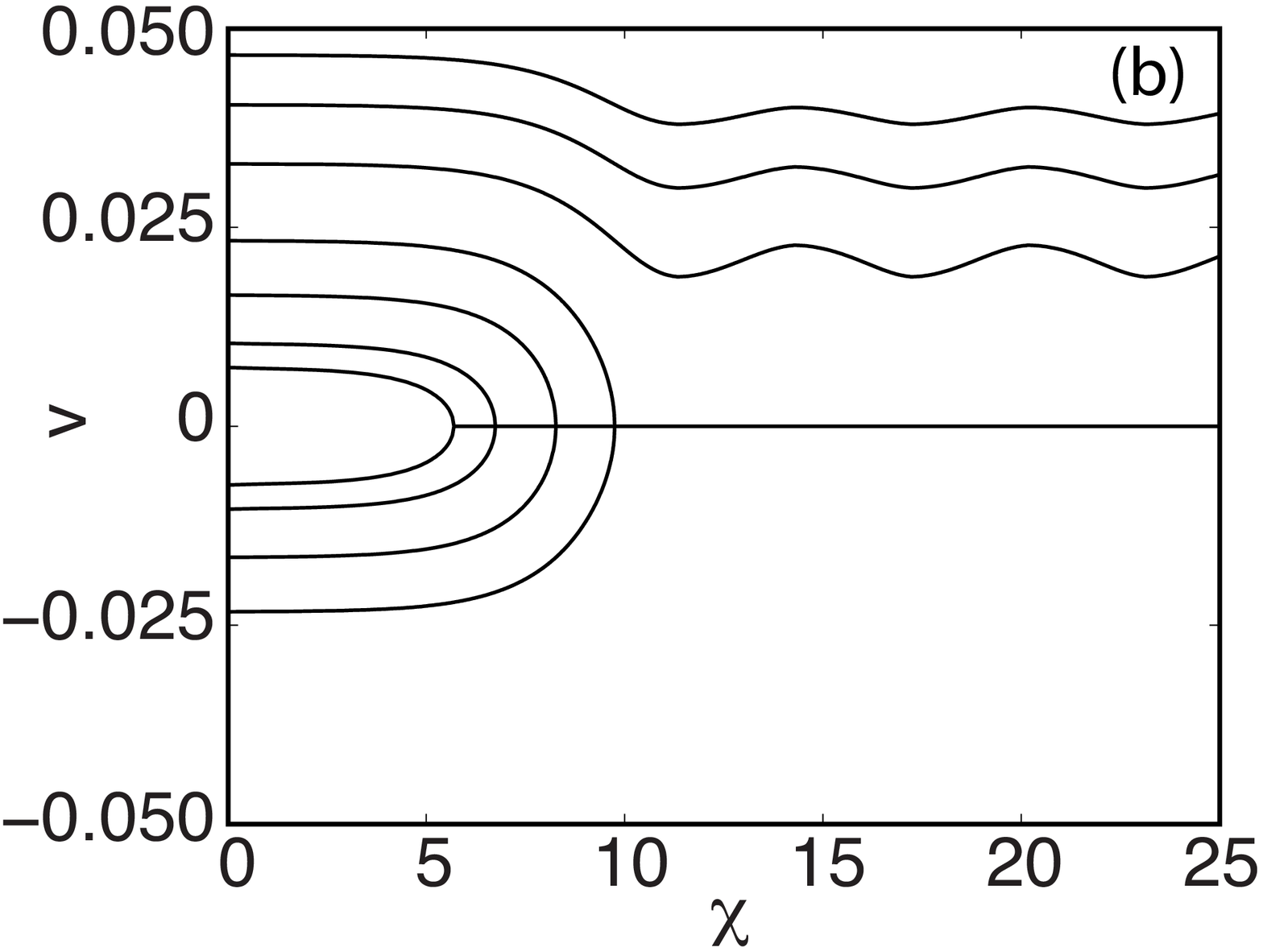}
\end{center}
\vspace{-24pt}
\caption{Electron (a) and ion (b) phase spaces for \(T_e = 10\) keV, \(M = 1.62\), \(\Theta = \Gamma = 1\) and \(T_i = 0.5\) keV.}\label{fig9}
\end{figure}

\begin{figure}[ht!]
\begin{center}
\includegraphics[width=6.8cm]{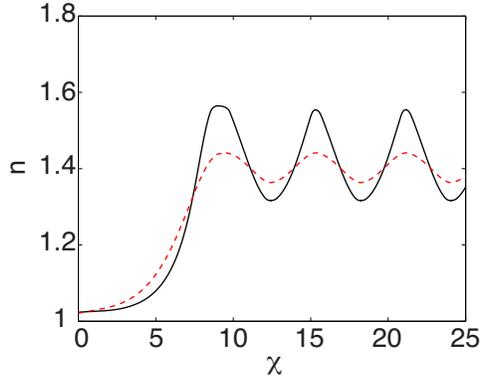}
\end{center}
\vspace{-24pt}
\caption{Spatial dependence of the ion (black, solid) and electron (red, dashed) density for \(T_e = 10\) keV, \(M = 1.62\), \(\Theta = \Gamma = 1\) and \(T_i = 0.5\) keV.}\label{fig10}
\end{figure}

Figure \ref{fig9} shows the electron and ion phase spaces, where the different populations (free, trapped, reflected) can be identified. The ion density follows the trend of the electrostatic potential \(\varphi\) (see figure \ref{fig10}). In the upstream region, the increasing potential decelerates and accumulates the ions which leads to an increase in the density. In the downstream, the ion density oscillates around a mean value.

%----------------------------------------------------------------------------
%----------------------------------------------------------------------------
%----------------------------------------------------------------------------
%----------------------------------------------------------------------------

%%%%%%%%%%%%%%%%%%%%%%%%%%%%%%%%%%%%%%%%%%%%%%%%%%%%%%%%%%%%%%%%%%%%
\section{Summary} \label{sec5}

In this paper, electrostatic shock solutions have been identified starting from an initial system of two colliding plasma slabs, containing two populations of hot electrons and cold ions and a population of trapped electrons. From the calculation of the pressure terms, the Sagdeev potential has been derived from the initial conditions for arbitrary density and temperature ratios, and a range for the steady state Mach number was presented for the stage where ion reflection starts to become important. For the first time, relativistic electron temperatures have been considered and approximations for the highly relativistic case have been presented. By introducing the steady state bulk velocity of the downstream population, an actual dependence of the Mach number on the initial density and temperature ratios was gained, bridging the range between the minimum and maximum Mach numbers. The critical Mach number at which ion reflection appears, is achieved by increasing the electron temperature in the upstream plasma, which will lead to an increase of the sound speed \(c_s\) and produce shock reflected ions with high energy.

In the context of applications of shock-accelerated ions, e.\,g.\ for medical purposes, it is important to achieve high energies and the realization of the theory is restricted by experimental feasibility. As shown in Fiuza et al.\ (2012) \cite{PRL}, for typical experimental conditions associated with laser-driven shocks, ion acceleration occurs close to the critical Mach number \(M \approx M_{cr}\), so that the reflected ions will have a velocity \(v_{i,refl} = 2 v_{sh} = 2 M_{cr} c_s\), which is determined by the condition for the critical Mach number. To simply reflect ions from the shock, figures \ref{fig:mmaxgamma} and \ref{CompSorasio}a show that large density ratios \(\Gamma\) and low temperature ratios \(\Theta\) are favorable, since low Mach number shocks are easier to drive.
In order to achieve high energy ions, large Mach numbers \(M_{cr}\) and/or large ion sound speeds \(c_s\) are needed. An increase in the Mach number can be gained by a high initial temperature ratio \(\Theta\), which is equivalent with increasing the energy of the slabs for a fixed upstream \(c_s\), and a high piston velocity, which provides a high momentum transfer to the plasma (i.\,e.\ initial relative fluid velocity between the two slabs) \cite{FS13b}. The sound speed can be increased by increasing the actual value of the electron temperature. This works well in near-critical density plasmas as a significant fraction of the laser energy can be absorbed by the plasma \cite{PRL}.

The theoretical model was extended by a population of reflected ions back into the upstream region, which transforms the solitary wave solution in the upstream region into a shock solution. The result is an oscillatory component in the downstream electrostatic potential. Our analysis shows that the general trends without ion reflection are valid, and that the inclusion of a reflected ion population leads to a slight increase in the Mach number. We note the similarity with shocks in quantum plasmas, where such a combination of dissipative and dispersive effects was found as well, although the underlying mechanisms are different \cite{BM08}.

%%%%%%%%%%%%%%%%%%%%%%%%%%%%%%%%%%%%%%%%%%%%%%%%%%%%%%%%%%%%%%%%%%%%

\acknowledgments
This work was partially supported by the European Research Council (ERC-2010-AdG Grant 267841) and FCT (Portugal) grants SFRH/BPD/65008/2009, SFRH/BD/38952/2007, and PTDC/FIS/111720/2009. We would like to thank Prof. G. Coppa and Prof. R. Bingham for fruitful discussions.

%%%%%%%%%%%%%%%%%%%%%%%%%%%%%%%%%%%%%%%%%%%%%%%%%%%%%%%%%%%%%%%%%%%%

\bibliography{IonShockAcc_theory_nr_v3}

\end{document}